# Phase stability determination of negative thermal expansion silicates by theory and experiment

Andreas Erlebach,[a] Ghada Belhadj Hassine,[a] Christian Thieme,[b] Katrin Thieme,[b]

Christian Rüssel,[c] Marek Sierka[a]*

[a] Otto-Schott-Institute for Materials Research, Friedrich Schiller University Jena, Löbdergraben 32, 07743 Jena, Germany

[b] Fraunhofer Institute for Microstructure of Materials and Systems IMWS, Walter-Hülse-Straße 1, 06120, Halle (Saale), Germany

[c] Otto-Schott-Institute for Materials Research, Friedrich Schiller University Jena, Fraunhoferstr. 6, 07743 Jena, Germany

**Abstract**

Materials that exhibit zero thermal expansion have numerous applications, ranging from everyday ceramic hobs to telescope mirrors to devices in optics and micromechanics. These materials include glass ceramics containing crystal phases with negative thermal expansion in at least one crystallographic direction, such as $Ba_{1-x}Sr_xZn_{2-2y}Mg_{2y}Si_2O_7$ solid solutions. However, the volume increase associated with the martensitic phase transformation in these crystals often hinders their use as zero thermal expansion materials at operating temperatures near the transition temperature $T_t$. Here, an approach to rapidly predict $T_t$ of such materials as a function of chemical composition based on a combination of density functional theory simulations and experiments has been developed and applied to $Ba_{1-x}Sr_xZn_{2-2y}Mg_{2y}Si_2O_7$. Its central element is the modeling of free energy as a function of temperature and chemical composition using a composition-dependent Debye model augmented by an empirical correction, which incorporates the effects of anharmonic lattice vibrations. This approach provides $T_t$ predictions with an estimated uncertainty of about $\pm 100$ K, which is similar to the accuracy of computationally much more demanding simulations of polymorphous phase transitions. In addition, this approach allows computationally efficient determination of the chemical compositions at which the $Ba_{1-x}Sr_xZn_{2-2y}Mg_{2y}Si_2O_7$ phase with the desired thermal properties will be formed during synthesis, facilitating the targeted design of zero thermal expansion materials.

1. **Introduction**

Materials showing zero thermal expansion (ZTE) have numerous applications ranging from commonplace ceramic hobs through telescope mirrors to devices in micromechanics.[1–3] Among these materials are glass ceramics containing crystal phases with negative thermal expansion (NTE) in at least one crystallographic direction,[4] such as the high temperature (HT) crystal phase of $Ba_{1-x}Sr_xZn_{2-2y}Mg_{2y}Si_2O_7$ solid solutions (BZS). The pronounced composition dependence of the thermal expansion coefficient of BZS allows straightforward tailoring of the thermomechanical properties of the corresponding glass ceramics.[5,6] However, the martensitic phase transition from the low temperature (LT) to the HT phase of BZS is connected with an increase in volume of about 3%.[7] Crack formation resulting from such a large volume change hampers exploitation of the tunable NTE of the HT BZS phase and its application as ZTE material at operating temperatures close to the phase transition temperature $T_t$. Therefore, the precise knowledge of $T_t$ is indispensable for the targeted design of BZS glass ceramics. It was demonstrated, that the incorporation of $Mg^{2+}$ and $Sr^{2+}$ into the crystal lattice of $BaZn_2Si_2O_7$ shows strongly opposite effects on $T_t$.[5,8] While the total substitution of $Zn^{2+}$ with $Mg^{2+}$ in $BaZn_{2-2y}Mg_{2y}Si_2O_7$ increases $T_t$ from about 550 K ($y = 0$) to 1210 K ($y = 1$),[5] the incorporation of $Sr^{2+}$ in $Ba_{1-x}Sr_xZn_2Si_2O_7$ solid solutions considerably decreases $T_t$ down to 473 K even at $Sr^{2+}$ concentrations as low as $x = 0.06$.[8] In addition, crystallization of glasses with ($x = 0.06$, $y = 0$) yields mixtures of the HT and LT phase. If the $Sr^{2+}$ concentration is further increased ($x > 0.1$), only the HT phase emerges during glass crystallization. This indicates that the substitution of $Ba^{2+}$ with $Sr^{2+}$ leads to thermodynamic stabilization of the HT phase substantially lowering $T_t$. At $x > 0.1$ and $T_t$ below 470 K, the martensitic phase transition is kinetically hindered and hence $T_t$ could not be observed experimentally between 100 and 900 K.[8] Owing to the opposite effects of $Mg^{2+}$ and $Sr^{2+}$ substitutions, the incorporation of both $Mg^{2+}$ and $Sr^{2+}$ into $BaZn_2Si_2O_7$ provides a simple way to tune the phase stability of the HT phase and $T_t$ over a large temperature range from (theoretical) 0 up to 1200 K.

However, the experimental characterization of the high number of possible chemical compositions of $Ba_{1-x}Sr_xZn_{2-2y}Mg_{2y}Si_2O_7$ would be a very time-consuming task.

For this reason, the prediction of $T_t$ using atomistic simulations would facilitate efficient targeted design of these BZS based ZTE glass ceramics. In particular, simulations at the density functional theory (DFT) level proved successful for prediction of polymorphic phase transitions.[9,10] Predictions of the phase stability of crystalline polymorphs require the calculation of free energy as a function of temperature. For this purpose, the vibrational density of states (VDOS) and the corresponding vibrational free energy can be calculated, *e.g.*, employing the harmonic approximation (HA).[10] Within the HA, however, the vibrational frequencies are assumed to be independent of volume and temperature. Therefore, predictions using the HA are limited in accuracy, in particular at elevated temperatures and for materials showing strongly anharmonic lattice vibrations.[11] More accurate, but still qualitative predictions of the phase stability are possible by combining DFT simulations with the quasiharmonic approximation (QHA), which takes the volume dependence of the VDOS into account (*e.g.*, ref. [12]). Quantitative predictions of phase stabilities require explicit consideration of anharmonicity at a given temperature. This is most commonly achieved using *ab initio* molecular dynamics (MD) simulations.[13–15] However, applying *ab initio* MD or phonon calculations using the QHA to a large number of chemical compositions is computationally very demanding.

In this work, an approach is developed for rapid predictions of the martensitic phase transition temperature that employs a combination of simulations at the DFT level and experiments. It is applied to $Ba_{1-x}Sr_xZn_{2-2y}Mg_{2y}Si_2O_7$ solid solutions. For this, composition dependent model functions for the VDOS are derived for the solid solution series of $Ba_{1-x}Sr_xZn_2Si_2O_7$ ($y = 0$) and $BaZn_{2-2y}Mg_{2y}Si_2O_7$ ($x = 0$) using phonon calculations along with the HA. Subsequently, an empirical correction is applied to the VDOS for consideration of anharmonic effects employing experimentally determined $T_t$. The resulting thermodynamic model is then used for predictions of $T_t$ of $Ba_{1-x}Sr_xZn_{2-2y}Mg_{2y}Si_2O_7$ ($x, y \geq 0$) solid solutions.

## 2. Computational details

All DFT calculations employed the *Vienna ab initio Simulation Package* (VASP).[16,17] The PBEsol exchange correlation functional[18,19] in combination with the empirical dispersion correction of Grimme et al.[20] (D3) was used along with the Projector Augmented Wave (PAW) method.[21,22] All simulations used the standard PAW potentials for all elements with two, four, six, ten, and twelve valence electrons for Mg, Si, O, Sr/Ba, and Zn, respectively. Calculations under constant pressure and constant volume conditions were performed with an energy cutoff of 900 and 400 eV for the plane wave basis sets, respectively. Integration of the first Brillouin zone used Monkhorst-Pack[23] grids with a linear $k$-point density of about 13 Å for each reciprocal lattice vector. Finally, calculation of VDOS for the structure models of the $Ba_{1-x}Sr_xZn_2Si_2O_7$ ($y = 0$) and $BaZn_{2-2y}Mg_{2y}Si_2O_7$ ($x = 0$) solid solution series were performed by employing the finite difference (frozen-phonon) approach implemented in the program Phonopy.[24] For this, 2×1×1 and 2×1×2 supercells for the LT and HT phase were used, respectively. Thermodynamic quantities were calculated from phonon frequencies[24,25] by employing Monkhorst-Pack grids with a linear $k$-point density of about 140 1/Å$^{-1}$ for each reciprocal lattice vector. The $k$-point grids used for all unit cells are summarized in Table S2 (see supporting information).

Simulations at the DFT level were applied to the LT and HT phase of $Ba_{1-x}Sr_xZn_{2-2y}Mg_{2y}Si_2O_7$ solid solutions for 25 chemical compositions ranging from BaZn ($x, y = 0$) to SrZn ($x = 1, y = 0$),

BaMg ($x = 0$, $y = 1$) as well as SrMg ($x = 1$, $y = 1$). In addition, structure models were generated for intermediate compositions in 0.25 steps for $x$, $y$ denoted as Sr(100$x$)Mg(100$y$), *e.g.*, Sr25Mg50 and Sr25Zn refer to the chemical composition ($x = 0.25$, $y = 0.5$) and ($x = 0.25$, $y = 0.0$), respectively.

Starting from the crystal structure of BaZn (LT and HT phase),[7] initial structure models for BaMg, SrZn and SrMg were constructed by replacing the corresponding ions in their lattice positions. In order to locate the lowest energy structures of $Ba_{1-x}Sr_xZn_2Si_2O_7$ ($y = 0$) and $BaZn_{2-2y}Mg_{2y}Si_2O_7$ ($x = 0$) solid solutions, initial structure models were generated by employing the crystal structure of the LT and HT phase of BaZn along with the side order-disorder (SOD) program.[26] SOD generates all possible structure models using the crystal symmetry of BaZn LT (space group *C*2/*c*, No. 15) and BaZn HT (space group *Cmcm*, No. 63), respectively, and replaces $Ba^{2+}$ and $Zn^{2+}$ on every symmetry inequivalent lattice site. Next, single point calculations were applied to every structure model generated. Then, all structures of the HT phase were geometrically optimized using constant (zero) pressure conditions. Due to the high number of structure models in case of the LT phase of BaMg25, BaMg50 and BaMg75 (up to 1742 for BaMg50), the configurations were grouped into sets of structure models having the same crystal symmetry (space group). Then, constant (zero) pressure geometry optimizations were applied to structure models selected from each set with relative energies of less than 1 kJ/mol with respect to the lowest energy structure in the corresponding set (about 40 optimizations per composition). Finally, the initial structure models for $Ba_{1-x}Sr_xZn_{2-2y}Mg_{2y}Si_2O_7$ with both $x > 0$ and $y > 0$, were generated using the combination of the $Ba^{2+}/Sr^{2+}$ and $Zn^{2+}/Mg^{2+}$ positions of the lowest energy structures obtained for the solid solutions ($x = 0$, $y > 0$) and ($x > 0$, $y = 0$).

## 3. Theory

Central quantity for phase stability predictions in the present work is the Helmholtz free energy $F(T)$ as a function of temperature $T$, which can be separated into the contributions

$$F(T) = E_0 + E_{\text{ZPE}} + F_{\text{vib}}(T) - TS_{\text{conf}}, \tag{1}$$

with the total lattice energy $E_0$ at 0 K, the zero-point vibrational energy $E_{\text{ZPE}}$ and configurational entropy $S_{\text{conf}}$. The calculation of lattice energies $E_{0,i}$ for all atomic configurations $i$ allows to determine the configurational partition function $Q = \sum_i \exp(-\beta E_{0,i})$ and the probability $P_i = \exp(-\beta E_{0,i}) Q^{-1}$ for the occurrence of a configuration $i$ in thermodynamic equilibrium ($\beta = (kT)^{-1}$).[26,27] This allows average properties $\overline{X}$ such as lattice energies or cell parameters to be calculated as $\overline{X} = \sum_i P_i X_i$ using with the quantities $X_i$ determined for each configuration $i$. In addition, the configurational entropy is given as

$$S_{\text{conf}} = -k \sum_i P_i \ln P_i. \tag{2}$$

The vibrational free energy $F_{\text{vib}}(T)$ can be calculated from the vibrational entropy $S_{\text{vib}}$ using[11,28]

$$F_{\text{vib}}(T) = -\int_0^T S_{\text{vib}}(T')\, dT'. \tag{3}$$

$S_{\text{vib}}$ of solids can be calculated using the vibrational entropy of a single harmonic oscillator (Einstein model) with frequency $\nu$ and the VDOS $g(\nu)$[11]

$$S_{\text{vib}}(T) = k \int_0^{\nu_{\max}} \left( \frac{z}{\exp(z)-1} - \ln(1-\exp(-z)) \right) g(\nu)\, d\nu, \tag{4}$$

with $z = \frac{h\nu}{kT}$ and the cutoff frequency $\nu_{\max}$. In this work, the harmonic Debye model for anisotropic solids is employed as approximate VDOS using three Debye temperatures $\theta_{i0}$[29]

$$g(\nu) = 9N \sum_{i=1}^{3} \left(\frac{h}{k\theta_{i0}}\right)^3 \nu^2. \tag{5}$$

For both LT and HT phases of BZS investigated in this work the Debye temperatures $\theta_{i0}(x,y)$ are assumed to depend linearly on the chemical composition $(x,y)$

$$\theta_{i0}(x,y) = \theta_{\text{BaZn},i} + B_1 x + B_2 y, \tag{6}$$

along with the adjustable parameters $\Theta_{\text{BaZn},i}$, $B_1$, and $B_2$.[30] Equations (3) to (6) define a model function of the harmonic $F_{\text{vib}}(T)$ that also depends on the chemical composition. To obtain the required model parameters ($\Theta_{\text{BaZn},i}$, $B_1$, $B_2$), $F_{\text{vib}}(T)$ was fitted using the vibrational free energy (and $E_{\text{ZPE}}$) from the phonon calculations for different compositions as described above. The key assumption of this approach is the formation of solid solutions, *i.e.*, assuming minor, systematic changes of the structure and vibrational properties with varying composition (*cf.*, Figure 3 and Table 1).

In case of the HA, $\theta_{i0}$ are independent of $T$. A correction of $S_{\text{vib}}$ for consideration of anharmonic effects can be obtained employing temperature dependent vibrational frequencies, *i.e.*, defining temperature dependent $\theta_i(T)$[11,28,31,32]

$$\theta_i(T) = \theta_{i0} e^{aT}. \tag{7}$$

Similar to $\theta_{i0}$, the anharmonicity parameter $a$ is assumed to depend linearly on the chemical composition using the adjustable parameters $a_0$, $C_1$, and $C_2$

$$a(x,y) = a_0 + C_1 x + C_2 y. \tag{8}$$

As shown in Figure 1, this correction is applied to $F_{\text{vib}}$ of the HT phase so that the free energy difference $\Delta F(T_t) = F_{\text{HT}}(T_t) - F_{\text{LT}}(T_t) = 0$ at the experimentally observed phase transition temperature $T_t$. Derivation of the required parameters used experimental data for the series Ba$_{1-x}$Sr$_x$Zn$_2$Si$_2$O$_7$ ($y = 0$)[8] and BaZn$_{2-2y}$Mg$_{2y}$Si$_2$O$_7$ ($x = 0$).[5]

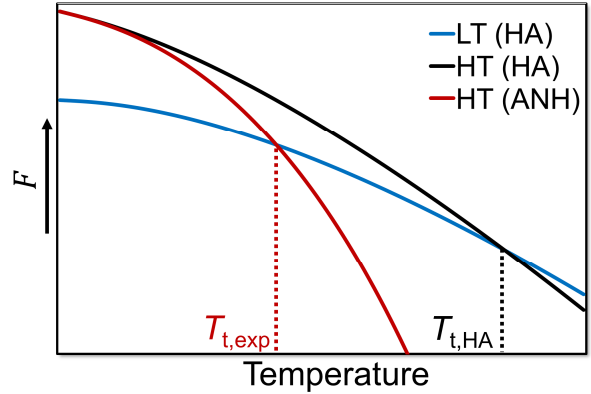

**Figure 1.** Empirical correction of the harmonic (HA) free energy $F$ (*cf.* eq 1) of the HT phase (yielding $T_{t,HA}$) using experimentally observed transition temperatures $T_{t,exp}$ for consideration of anharmonic (ANH) effects.

Finally, in order to obtain a continuous function for relative energies of the HT with respect to LT phase $\Delta E_0$ (at 0 K), a quartic polynomial was fitted to the lattice energy differences of lowest energy structures calculated using DFT simulations described in Section 2. This allows modeling of $F(T)$ and prediction of $T_t$ for the complete composition range ($x$, $y$) of $Ba_{1-x}Sr_xZn_{2-2y}Mg_{2y}Si_2O_7$ solid solutions.

## 4. Experimental procedure

In order to validate the calculated phase transition temperatures, some compositions, which behavior was up to now not reported in the literature were synthesized via conventional ceramic route. Therefore, the respective raw materials $BaCO_3$, $SrCO_3$, ZnO, MgO, and $SiO_2$ were carefully homogenized and afterwards heat treated at 1000 °C for 5 h. Then, the obtained materials were powdered and again heat treated at 1100 °C for 5 h and 2 times at 1200 °C for 5 h. In between the heat treatments, the samples were crushed and powdered. At the end of the synthesis procedure, each sample was again crushed and milled to a fine powder. The repetition of the milling procedure leads to highly homogeneous powders. The phase purity was afterwards checked via powder X-ray diffraction (XRD).

For those measurements, a Rigaku MiniFlex 300 with CuK$_\alpha$ radiation and Bragg-Brentano geometry was used. The diffraction patterns were recorded in the 2θ range from 10 to 60°. The patterns were compared with the theoretical diffraction lines of $Ba_{0.6}Sr_{0.4}Zn_2Si_2O_7$ (ICSD 429938). The phase transition temperatures were measured via differential scanning calorimetry (DSC) and dilatometry. In the case of DSC measurements, 60 mg of powder were heated up to 1200 °C in $Al_2O_3$-crucibles with a rate of 10 K/min using a Linseis DSC Pt-1600. Some compositions lead to inconclusive results which made it necessary to determine the phase transition temperature additionally with dilatometry. For this purpose, a Linseis DIL L75VD dilatometer equipped with an $Al_2O_3$ measuring system was used. The samples were prepared by filling an appropriate amount of fine powders of the respective samples into an $Al_2O_3$ boat with a length of around 20 mm. The powder was slightly compressed with a spatula and then given into a conventional muffle furnace with a temperature accuracy of ± 5 K, where they were heat treated at 1200 °C for 10 h. This heat treatment led to sintering of the powders so that compact samples could be obtained. The ends of the samples were ground so that two parallel planes were obtained. Those samples were transferred into the dilatometer, where they were heated with 2 K/min. The thermal expansion behavior and especially the coefficient of thermal expansion (CTE) cannot accurately be measured from such samples, because of the microstructure after the sintering process, which perhaps contains large and numerous cracks. Such cracks strongly affect the thermal expansion. Nevertheless, the phase transition temperature can exactly be determined.

## 5. Results and discussion

### 5.1. Atomistic modeling of BZS solid solutions

Figure 2 shows the calculated lowest energy structures for the LT and HT phases of $Ba_{0.5}Sr_{0.5}Zn_2Si_2O_7$ (Sr50Zn) and $BaZnMgSi_2O_7$ (BaMg50). Comparison of cell parameters and relative energies $\Delta E_0$ are summarized in Table 1. In addition, the dependence of cell parameters $a$ and $c$ for $Ba_{1-x}Sr_xZn_2Si_2O_7$ on the chemical composition obtained from DFT simulations and experiments[8] are depicted in Figure 3a.

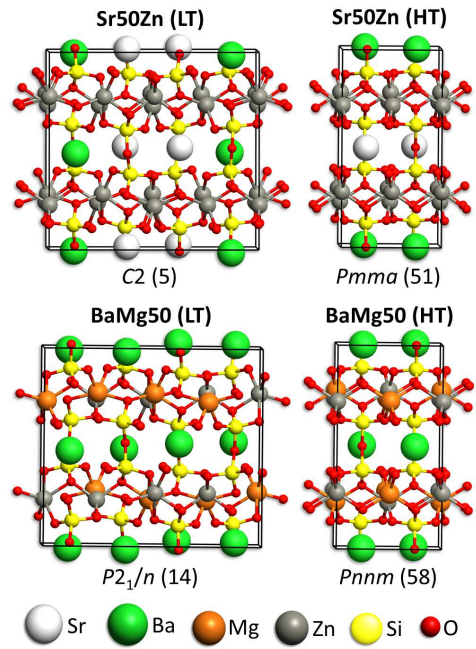

**Figure 2.** Lowest energy structures of the low (LT) and high temperature (HT) phase for $Ba_{0.5}Sr_{0.5}Zn_2Si_2O_7$ (Sr50Zn) and $BaZnMgSi_2O_7$ (BaMg50) obtained from DFT simulations.

For all chemical compositions, a monoclinic crystal structure for the LT phase and an orthorhombic structure for the HT phase was obtained. The fundamental assumption for construction of the atomistic models is the formation of solid solutions, that is the basic crystal structure of $BaZn_2Si_2O_7$ (BaZn) remains unchanged by substitution of $Ba^{2+}$ with $Sr^{2+}$ and $Zn^{2+}$ with $Mg^{2+}$ (Figure 2), respectively. Except for the crystallographic $c$-axis of the LT phase for BaZn$_{2-2y}$Mg$_{2y}$Si$_2$O$_7$ solid solutions, the lattice parameters change only slightly with the chemical composition. In addition, the cell parameters *a* and *c* vary almost linearly for the HT phase of Ba$_{1-x}$Sr$_x$Zn$_2$Si$_2$O$_7$, which is in good agreement with experimental observations (Figure 3a). For SrZn (*x* = 1, *y* = 0), no experimental values are available since neither the LT nor the HT phase was obtained experimentally.[8] Similar linear dependences of the cell parameters on the chemical composition were obtained for BaZn$_{2-2y}$Mg$_{2y}$Si$_2$O$_7$ solid solutions from experiments[5,6] in excellent agreement (deviation less than 3%) with DFT calculated lattice parameters (*cf.* supporting information, Table S1). Such linear change of lattice parameters of BaZn caused by substitution of Ba$^{2+}$ and Zn$^{2+}$ supports, according to Vegard's law,[33] the assumption that indeed solid solutions as depicted in Figure 2 are formed.

**Table 1.** Change of lattice parameters [Å] with respect to BaZn, space groups, and relative lattice energies $\Delta E_0$ [kJ/mol] of the HT with respect to the LT phase, obtained from DFT calculations.

| Sample | | Lattice parameters | | | Space group | $\Delta E_0$ |
|---|---|---|---|---|---|---|
| | | *a* | *b* | *c* | | |
| BaZn | LT | 7.181 | 12.691 | 13.680 | *C*2/*c* (15) | 3.9 |
| | HT | 7.802 | 12.956 | 6.614 | *Cmcm* (63) | |
| BaMg50 | LT | -0.035 | -0.088 | 0.087 | *P*2$_1$/*c* (14) | 17.0 |
| | HT | -0.042 | 0.025 | 0.027 | *Pnnm* (58) | |
| Sr50Zn | LT | -0.051 | 0.036 | -0.257 | *C*2 (5) | 3.1 |
| | HT | -0.085 | -0.018 | -0.086 | *Pmma* (51) | |
| Sr50Mg50 | LT | -0.081 | -0.090 | -0.162 | *P*2$_1$ (4) | 14.7 |
| | HT | -0.093 | -0.005 | -0.073 | *Pm* (6) | |
| Sr25Mg75 | LT | -0.051 | -0.112 | -0.046 | *P*1 | 22.7 |
| | HT | -0.083 | 0.012 | -0.003 | *P*1 | |

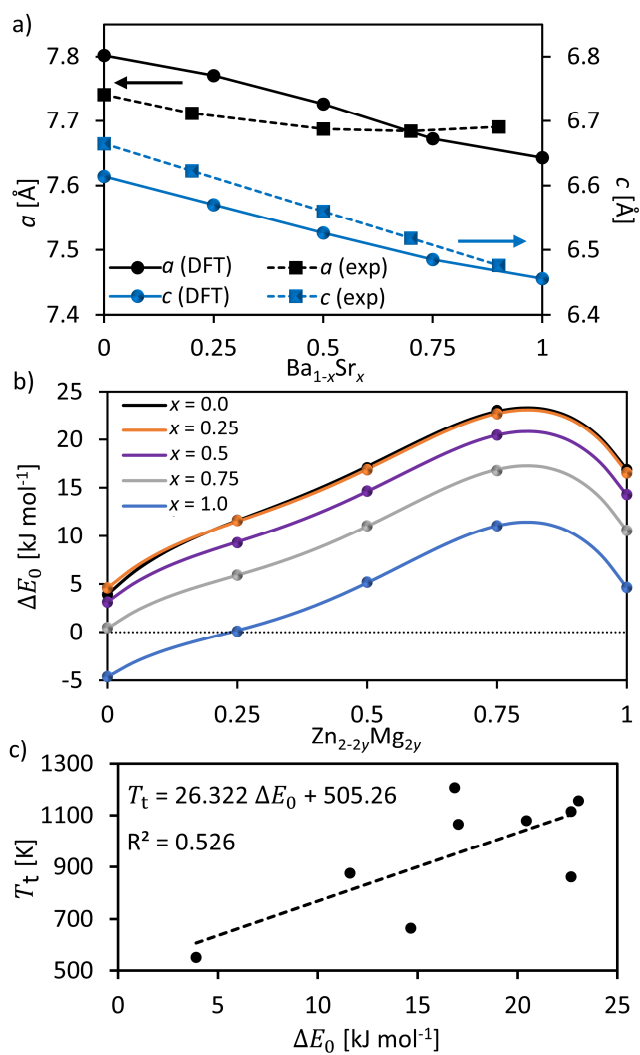

**Figure 3.** Change of a) lattice parameters $a$ and $c$ of the HT phase for $Ba_{1-x}Sr_xZn_2Si_2O_7$ calculated using DFT in comparison with experimental (exp) values,[8] b) fit of lattice energies $\Delta E_0$ of the HT with respect to the LT phase as a function of chemical composition $(x, y)$, and c) qualitative correlation of calculated $\Delta E_0$ with experimentally observed $T_t$.

The fit of relative energies $\Delta E_0$ as a function of $(x, y)$ to a quartic polynomial is depicted in Figure 3b showing very good agreement with DFT calculated values along with a mean absolute error (MAE) of 0.1 kJ/mol. For $\Delta E_0 \leq 0$ the HT phase is predicted to be thermodynamically stable for all temperatures and, thus, $T_t$ is set to 0 K. For $Sr^{2+}$ concentrations between $x = 0$ and 0.25, $\Delta E_0$ remains virtually unchanged, yet considerably decreases for larger $x$. By contrast, $\Delta E_0$ increases with increasing $Mg^{2+}$ concentration $y$, except for the maximum at $y = 0.75$, which is about 6 kJ/mol higher compared to $y = 1$. Such deviation is expected to be similar to the accuracy of DFT simulations. However, there is a qualitative correlation of $\Delta E_0$ with experimentally observed $T_t$ (Figure 3c),[5,6,8] indicating that the calculation of $\Delta E_0$ for other chemical compositions can indeed be used for prediction of $T_t$.

In case of $Ba_{1-x}Sr_xZn_2Si_2O_7$ solid solutions, all constructed structure models were geometrically optimized. This allows the calculation of the average cell parameters as well as the average relative energy $\overline{\Delta E_0}$. The differences between the average cell parameters and those calculated for the lowest energy structures are smaller than 0.03 Å. Similarly, the deviation between $\Delta E_0$ of the lowest energy configurations and $\overline{\Delta E_0}$ are lower than 1 kJ/mol for all chemical compositions $x$. Due to these minor deviations, it can be assumed that $\Delta E_0$ calculated for the lowest energy structures can be used for the thermodynamic modeling of $T_t$ as a function of the chemical composition. Because of the high number of possible atomic configurations for the LT phase of $BaZn_{2-2y}Mg_{2y}Si_2O_7$ solid solutions, geometry optimizations were only applied to a subset of structure models. Therefore, the lattice energies obtained from single point calculations of the generated structure models (*cf.* eq 3) are used for estimation of the configurational entropy difference $\Delta S_{conf} = S_{conf,HT} - S_{conf,LT}$. Assuming the glass crystallization temperature of 1200 K,[34] the free energy contributions $-T\Delta S_{conf}$ are only -1 kJ/mol (per formula unit) for $Ba_{1-x}Sr_xZn_2Si_2O_7$ and 2 kJ/mol for $BaZn_{2-2y}Mg_{2y}Si_2O_7$ solid solutions, respectively. These small free energy differences, which are virtually independent of the chemical compositions ($x$, $y$), can be rationalized with the similar LT and HT crystal structures that differ only in crystal symmetry (*cf.*

Table 1). Since such deviations in free energy are expected to be lower than DFT accuracy for calculation of $\Delta E_0$, $\Delta S_{\text{conf}}$ is assumed to be zero. Furthermore, the empirical correction of the harmonic vibrational free energy considers not only the effects of anharmonic lattice vibrations on $T_t$ but also includes the DFT error of $\Delta E_0$ as well as the neglect of $\Delta S_{\text{conf}}$.

**5.2. Modeling of vibrational free energy**

Table 2 shows the fitted parameters for modeling of the composition dependent Debye temperatures $\theta_{i0}(x, y)$ (eq 6) and the composition dependent anharmonicity parameter $a(x, y)$ (eq 8). In addition, Figure 4a compares the harmonic vibrational density of states (VDOS) determined using phonon calculations at the DFT level with the approximated Debye VDOS (*cf.* eq 5) for the HT phase of BaZn and BaMg. The Debye VDOS used the harmonic model for anisotropic solids with three Debye temperatures (representing the three spatial directions) resulting in the three parabolic curves shown in Fig. 4a (*cf.* 5). Parametrization of the Debey VDOS used a fit of $F_{\text{vib}}(T)$ (eqs 3-6) to the harmonic free energies obtained from DFT simulations for BaZn, BaMg50, BaMg, Sr50Zn, and SrZn.[29,30] The fitted free energies for the HT phases of BaZn and BaMg are shown in Figure S1 (see supporting information) as a representative example. In all cases, the vibrational free energies calculated using the Debye model are in very good agreement with the DFT results along with a MAE of 1.3 kJ/mol. Such minor free energy deviations are similar to DFT accuracy and translate to a $T_t$ prediction accuracy of about ±100 K (see Section 5.3).[10] Please note that the actual shape of the approximated VDOS $g(v)$ is less relevant rather than its integral shown in eq 4 used for calculation of $F_{\text{vib}}(T)$ (eq 3). However, the corresponding $T_t$ predicted by the HA at the DFT level disagree even qualitatively with the experimentally observed values, *e.g.*, for BaZn and BaMg the HA predicted $T_t$ of 2070 and 1830 K deviate by over 1500 K from the experimentally observed values of 553 and 1207 K, respectively.[5] Therefore, consideration of the effects of anharmonic lattice vibrations on the vibrational free energy $F_{\text{vib}}$ is indispensable for reliable predictions of $T_t$ over a wide composition range.

**Table 2.** Model parameters $\theta_{\text{BaZn},i}$, $B_1$, $B_2$ [K] of the composition dependent Debye temperatures (*cf.* eq 6) and anharmonicity parameters $a_0$, $C_1$, $C_2$ [K$^{-1}$] (*cf.* eq 8).

|    | $\theta_{\text{BaZn},1}$ | $\theta_{\text{BaZn},2}$ | $\theta_{\text{BaZn},3}$ | $B_1$ | $B_2$ | $a_0$ | $C_1$ | $C_2$ |
|----|------|------|--------|-------|-------|-------------------|------------------|-------------------|
| LT | 216.0 | 734.7 | 1399.2 | 56.14 | 8.88  | -                 | -                | -                 |
| HT | 198.9 | 837.2 | 1328.0 | 54.1  | 10.50 | $-1.356\times10^{-4}$ | $4.192\times10^{-5}$ | $-6.456\times10^{-4}$ |

Figure 4b shows $T_t$ calculated using the temperature dependent Debye model (*cf.* eq 7) for consideration of anharmonic effects on $F_{\text{vib}}$ of the HT phase in comparison with experimentally determined values used for model derivation. In case of the substitution of Ba$^{2+}$ by Sr$^{2+}$ ($x > 0$, $y = 0$), only two closely lying experimental values are available for $x = 0.02$ and $x = 0.06$ with $T_t = $ 543 K and 473 K, respectively. For the latter, a mixture of the LT and HT phase was obtained, while for higher Sr$^{2+}$ concentrations ($x > 0.1$) only the HT phase evolves during the synthesis and, consequently, no $T_t$ could be observed.[8] An atomistic model for such low Sr$^{2+}$ concentrations would require large unit cells and a vast number of calculations for localization of the lowest energy structure rendering simulations at the DFT level tremendously challenging. Therefore, values of the polynomial fit of $\Delta E_0$ (*cf.* Figure 3) were used for the fit of anharmonic model for $F_{\text{vib}}$ of the HT phase in this case. By contrast, five experimentally determined $T_t$ were used for model derivation in case of the ($x = 0$, $y > 0$) solid solution series.[5]

The largest deviation of $T_t$ calculated using the derived model, of about 100 K from the experimental value, is at $y = 0.75$, due to the DFT error for $\Delta E_0$ (see also Table 3). In order to estimate the uncertainty of the model predictions, ten fits were performed using every combination of two out of the five available experimental $T_t$ for ($x = 0$, $y > 0$). The prediction accuracy is about ±100 K and represented by the shaded area in Figure 4b. For comparison, an error in $\Delta E_0$ of only 0.8 kJ/mol, which is assumed to be lower than the DFT accuracy, translates into a deviation of a polymorphous phase transition temperature of about 90 K with respect to the experimentally obtained value.[10] That is, even if only two experimentally determined $T_t$ are available for calculation of the empirical correction, the deviation of the model predictions is in the same order of magnitude as the DFT accuracy. Therefore, the proposed procedure appears to be an efficient way to combine phonon calculations employing the HA with two experimental values for $T_t$ to include the effects of anharmonic lattice vibrations. This considerably reduces computational efforts compared to, *e.g.*, *ab initio* MD simulations. Moreover, as mentioned above, for BZS compositions with $x > 0.1$ and $y = 0$ only the HT phase emerges during synthesis and persists during cooling to room temperature. For this composition ($x = 0.1$, $y = 0$), the calculated $T_t$ is about 500 K indicated as dotted line in Figure 4b. Therefore, it can be assumed that for predicted $T_t$ below 500 K only the HT phase is obtained, while for $T_t > 500$ K only the LT phase is observed in crystalline samples at room temperature.

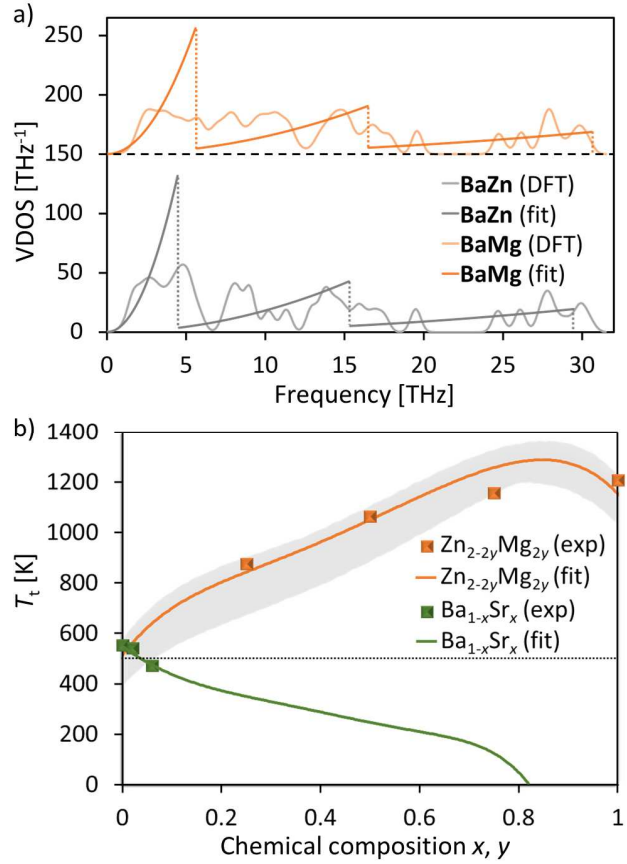

**Figure 4.** Modeling of a) the composition dependent vibrational density of states (VDOS) using the derived Debye model (eqs 5 and 6) of the HT phase for BaZn and BaMg compared to the DFT calculated VDOS and b) phase transition temperatures ($T_t$) as a function of the chemical composition ($x$, $y$) along with experimentally observed $T_t$ used for model parameterization.[5,8] The shaded area indicates the uncertainty of model predictions (see text). Below $T_t$ of approximately 500 K (dotted line in b) only the HT phase was obtained in experimental studies.

### 5.3. Predictions of the phase stability

Table 3 summarizes calculated and experimentally observed $T_t$ used for model parameterization[5,8] as well as $T_t$ for chemical compositions, which were not included in the least square fitting of the empirical correction, namely Sr25Mg50, Sr50Mg50, Sr25Mg75, and Sr50Mg75. In addition, Figure 5a shows the results of model predictions for $T_t$ as a function of the chemical composition of BZS solid solutions with $x$, $y$ between 0 and 1. $T_t$ values range from 0 K for compositions close to SrZn up to 1283 K for x = 0 and y = 0.82. The latter is overestimated due to error in $\Delta E_0$ (*cf.* Figures 3 and 4). For Sr50Mg50 and for Sr25Mg75, the predicted $T_t$ show very good agreement with the experimentally determined values, with differences of about 120 and 40 K, respectively. These deviations are approximately the same as the estimated uncertainty of ±100 K of the model predictions and in the same order of magnitude of the expected inaccuracy of DFT simulations.[10] This indicates that the assumption of linear composition dependencies of $\theta_{i0}(x,y)$ and $a(x,y)$ provide a good approximation for modeling of $F_{vib}$ of BZS solid solutions.

**Table 3.** Comparison of phase transition temperatures $T_t$ [K] obtained by calculations (calc) and experiment (exp) for different chemical compositions.

| Samples | $T_t$ (exp) | $T_t$ (calc) |
|---|---|---|
| BaZn | 550[a] | 495 |
| BaMg25 | 875[a] | 840 |
| BaMg50 | 1065[a] | 1048 |
| BaMg75 | 1156[a] | 1253 |
| BaMg | 1207[a] | 1142 |
| Sr25Mg50 | -[b] | 707 |
| Sr50Mg50 | 665 (DSC) | 547 |
| Sr25Mg75 | 861 (dil)<br>1114 (dil) | 821 |
| Sr50Mg75 | 1079 (dil) | 638 |

[a]taken from ref. 5
[b]no clearly identifiable effects in DSC or dilatometry (dil) curves

For Sr25Mg75 a second phase transition was observed at higher temperature (1114 K). In case of Sr50Mg75, only one phase transition at similarly high temperature (1079 K) was experimentally determined, which lies considerably above the predicted $T_t$ of 638 K. For Sr25Mg50 no phase transition was detected, neither by employing DSC nor dilatometry. This might be connected with too small enthalpy and structural changes caused by the martensitic phase transition. On the other hand, the phase transition could be suppressed by residual stresses in the microstructure induced during cooling of the samples.

Since the derived thermodynamic model for $F_{vib}$ assumes the formation of substitutional solid solutions, it considers only a continuous, linear dependence of the harmonic VDOS on the chemical composition. In addition, the model derivation included only the solid solution series for substitutions of either $Ba^{2+}$ by $Sr^{2+}$ ($x > 0$, $y = 0$) or $Zn^{2+}$ by $Mg^{2+}$ ($x = 0$, $y > 0$). That is, the model predictions assume that the linear composition dependence of the VDOS also holds for $Sr^{2+}/Mg^{2+}$ co-substitutions. These co-substitutions cause a linear change of the lattice parameters (*cf.* Ref. 6) similar to $Ba_{1-x}Sr_xZn_2Si_2O_7$ ($x > 0$, $y = 0$, *cf.* Fig. 3a) indicating the formation of substitutional solid solutions and supporting the model assumptions. Therefore, it is expected that the estimated $T_t$ prediction accuracy of ±100 K also applies to co-substitutions as proved by the experimental results for Sr50Mg50 and Sr25Mg75 (*cf.* Table 3). However, the proposed theoretical approach does neither consider a second phase transition at higher temperatures nor the potentially suppressed phase transition in case of Sr25Mg50. In addition, synthesis of BZS solid solutions yields defect containing crystallites showing local residual stresses and highly anisotropic microstructures[35] leading to the formation of cracks that also influence the macroscopic properties of the obtained material.[34] By contrast, the thermodynamic model assumed defect free, single crystalline and stress free BZS solid solutions. Therefore, the employed procedure provides predictions $T_t$ with an uncertainty of about $\pm 100$ K only if the influence of lattice defects and the microstructure on $T_t$ is not too pronounced, which could also rationalize the hindrance of the phase transition in case of Sr25Mg50.

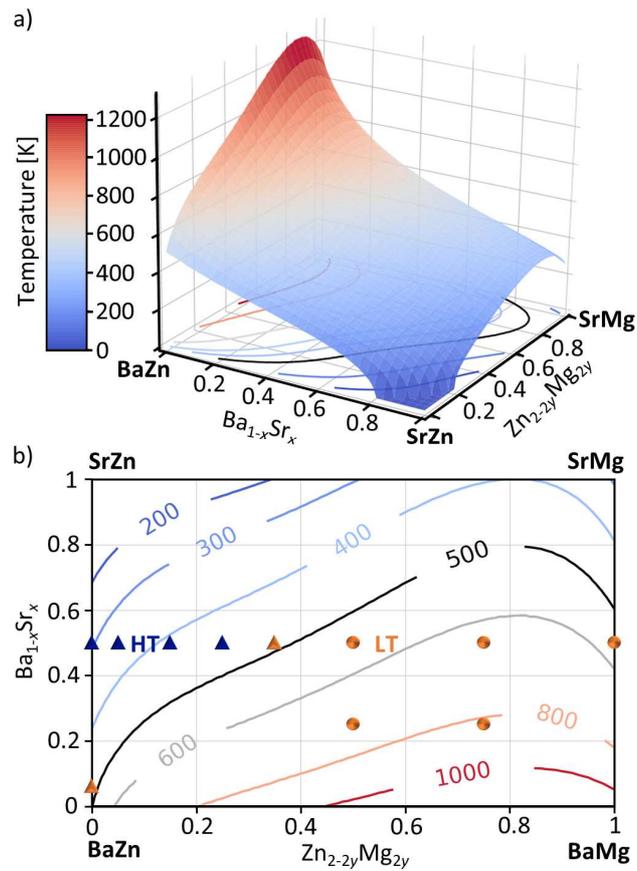

**Figure 5.** a) Plot of predicted phase transition temperatures (in K) as a function of the chemical composition. b) Contour lines of (a) along with compositions at which only the high temperature (HT) phase (blue triangles), only the low temperature (LT) phase (orange dots), and a mixture of both phases (orange triangle) was obtained in experiments.[6,8]

The contour plot of Figure 5a is shown in 5b along with the chemical compositions at which the LT (orange dots), HT (blue triangles) or mixtures of both phases (orange triangles) were obtained from solid state reaction, respectively. The empirical value for $T_t$ of 500 K under which solely the HT phase is stable (dashed line in Figure 4b) is also highlighted in Figure 5b. As mentioned above, for chemical compositions ($x > 0.1$, $y = 0$) only the HT phase is observed in the ceramic material, while a mixture of HT and LT was observed at ($x = 0.06$, $y = 0$). For the BZS solid solution series with $x = 0.5$ and $y > 0$, the samples contained only the HT phase up to $y = 0.25$. At $y = 0.35$, a mixture of LT and HT was observed along with a predicted $T_t$ of approximately 500 K. If the $Mg^{2+}$ concentration $y$ is further increased only the LT phase was found at room temperature. This applies also to the chemical compositions Sr25Mg50 and Sr25Mg75.

The considerable decrease of both, the $\Delta E_0$ calculated using DFT simulations and the experimentally observed $T_t$, with increasing $Sr^{2+}$ concentration $x$ clearly indicates that the HT is thermodynamically stabilized by substitution of $Ba^{2+}$ with $Sr^{2+}$. Most likely, the HT phase emerges at higher temperatures and transforms into the LT phase during cooling of the samples. For chemical compositions with predicted $T_t$ below 500 K, the martensitic phase transition is probably kinetically hindered such that only the HT phase is observed at room temperature. Both samples with predicted $T_t$ of about 500 K show a mixture of the LT and HT phase. This is probably connected with small, local fluctuations of the chemical composition such that only a certain fraction of the HT crystallites transforms into the LT phase during cooling of the sample. This allows the rapid localization of those chemical compositions at which the desired HT phase showing NTE can be obtained. Since the coefficients of thermal expansion are strongly anisotropic for both, the HT and the LT phase, also internal stresses formed during cooling might prevent a total conversion of the high into the low temperature phase during cooling. In addition, despite certain limitations mentioned above, the used computational procedure facilitates rapid prediction of the corresponding phase transition temperature. In this way, promising chemical compositions can be determined for further experimental characterization and more demanding simulations, such as *ab initio* MD simulations to provide deeper understanding of the structure-property relations of ZTE glass ceramics.

## 6. Conclusions

In summary, an approach for rapid prediction of the martensitic phase transition temperature $T_t$ based on a combination of DFT simulations and experiments has been developed and applied to $Ba_{1-x}Sr_xZn_{2-2y}Mg_{2y}Si_2O_7$ solid solutions. Central to this approach is the modeling of free energy as a function of temperature and chemical composition. For this, comprehensive sampling of atomic configurations was performed for localization of the lowest energy structures and calculation of the relative lattice energies. For BZS solid solution it has been found that the crystal structures and relative energies are almost independent of the precise arrangement of $Ba^{2+}$, $Sr^{2+}$, $Zn^{2+}$, and $Mg^{2+}$ ions on the respective lattice sites. Moreover, the contribution of configurational entropy to the free energy was found to be negligible. Next, a composition dependent Debye model was derived using the harmonic vibrational free energy obtained from phonon calculations at the DFT level.

However, predictions using the harmonic approximation were found not to agree even qualitatively with experimentally determined $T_t$. Therefore, an empirical correction of the free energy of vibration was proposed using experimentally determined $T_t$, which incorporates the effects of anharmonic lattice vibrations. In addition, this correction partially compensates for DFT errors and for the error of the simplified model used for VDOS. Including this correction yields model predictions of $T_t$ with an estimated uncertainty of about $\pm 100$ K, which is similar to DFT accuracy for simulations of polymorphous phase transitions. Predictions for chemical compositions not included in the model derivation show the same deviation from experimentally determined $T_t$.

Moreover, an empirical value for $T_t$ of 500 K was found at which ceramic materials yield only the HT phase, allowing rapid screening of chemical compositions. This is probably connected with kinetic hindrance of the martensitic phase transition. Thus, this approach, combining DFT simulations and experimental data, can be used for rapid predictions of both $T_t$ and the promising chemical composition for which the desired HT phase with NTE can be obtained.

ABBREVIATIONS

| | |
|---|---|
| HT | High temperature phase |
| LT | Low temperature phase |
| ZTE | Zero thermal expansion |
| NTE | Negative thermal expansion |
| CTE | Coefficient of thermal expansion |
| BZS | Barium zinc silicate solid solutions ($Ba_{1-x}Sr_xZn_{2-2y}Mg_{2y}Si_2O_7$) |
| HA | Harmonic approximation |
| QHA | Quasiharmonic approximation |
| MAE | Mean absolute error |
| ANH | Anharmonic |
| VDOS | Vibrational density of states |
| DSC | Differential scanning calorimetry |
| DFT | Density functional theory |
| MD | Molecular dynamics |

# AUTHOR INFORMATION

**Corresponding Author**

*Marek Sierka, email: marek.sierka@uni-jena.de

**Author Contributions**

The manuscript was written through contributions of all authors. All authors have given approval to the final version of the manuscript.

**Funding Sources**

This work was funded by the German Federal Ministry of Education and Research under the Grant Numbers 03VP01701, 03VP01702 and 13XP5122C.